\documentclass[aps,prl,twocolumn, showpacs,amsmath,amssymb,amsfonts,nofootinbib,long]{revtex4}
\usepackage{epsfig,latexsym}

\begin{document}

\title{Evolution of Magnetic Fields in Freely Decaying Magnetohydrodynamic Turbulence}

\author{Leonardo Campanelli$^{1,2}$}

\email{campanelli@fe.infn.it}

\affiliation{$^{1}${\it Dipartimento di Fisica, Universit\`a di Ferrara, I-44100 Ferrara, Italy
\\           $^{2}$INFN - Sezione di Ferrara, I-44100 Ferrara, Italy}}

\date{May, 2007}


\begin{abstract}
We study the evolution of magnetic fields in freely decaying
magnetohydrodynamic turbulence.
By quasi-linearizing the Navier-Stokes equation, we solve
analytically the induction equation in quasi-normal approximation.
We find that, if the magnetic field is not helical, the magnetic
energy and correlation length evolve in time respectively as $E_B
\propto t^{-2(1+p)/(3+p)}$ and $\xi_B \propto t^{2/(3+p)}$, where
$p$ is the index of initial power-law spectrum.
In the helical case, the magnetic helicity is an almost conserved
quantity and forces the magnetic energy and correlation length to
scale as $E_B \propto (\log t)^{1/3} t^{-2/3}$ and $\xi_B \propto
(\log t)^{-1/3} t^{2/3}$.
\end{abstract}


\maketitle


The origin of presently-observed large scale magnetic fields
throughout the universe is still unclear \cite{Magnetic}.
Essentially, there are two possible classes of mechanisms to
produce cosmic fields depending on when they are generated:
Astrophysical mechanisms acting during or after large-scale
structure formation, and mechanisms acting in the primordial
universe.
Magnetic fields created in the early universe (except those
generated during inflation), usually suffer from a ``small-scale
problem'', that is their comoving correlation length is much
smaller then the characteristic scale of the observed cosmic
fields. However, if magnetohydrodynamic (MHD) turbulence operates
during their evolution, an enhancement of correlation length can
occur, especially if the magnetic field is helical.
As pointed out by Banerjee and Jedamzik~\cite{Banerjee}, the
evolution of a magnetic field in the early universe goes through
different phases depending on the particular conditions of the
primordial plasma. In this paper, we are interested in the case of
magnetic fields evolving in the turbulent primordial universe well
before recombination epoch and when kinematic dissipative effects
are due to diffusing particles. Therefore, we are concerned with
the so-called phase of ``turbulent MHD''. In other phases, such as
``viscous MHD'' and ``MHD with ambipolar diffusion'' described in
Ref.~\cite{Banerjee}, the dynamics of the magnetic field is very
different from that studied here.
\\
The problem of determining the evolution properties of magnetic
fields in MHD turbulence has been deeply and widely discussed in
the literature using different methods and approximations.
A direct integration of the full set of MHD equations would allow
us to deeply understand the dynamics of freely decaying MHD
turbulence. However, MHD equations are quite difficult to handle
due to their high non-linearity and it has not been yet brought in
a definitive verdict for the evolution laws of magnetic energy and
correlation length (for recent numerical studies of freely
decaying magnetohydrodynamic turbulence see, e.g.,
Ref.~\cite{Biskamp1,Christensson,Banerjee}).


The turbulent MHD equations for incompressible fluids, in the case
of non-expanding universe, are \cite{Biskamp}:
\vspace{-0.1cm}
\begin{eqnarray}
\label{Eq1} && \partial_t {\textbf v} + {\textbf v} \cdot \nabla
{\textbf v} + \nabla p - \nu \nabla^2 {\textbf v} \, = \, {\textbf
J} \times {\textbf B},
\\
\label{Eq2} && \partial_t {\textbf B} \, = \, \nabla \times
({\textbf v} \times {\textbf B}) + \eta \nabla^2 {\textbf B},
\end{eqnarray}
and $\nabla \cdot {\textbf v} = \nabla \cdot {\textbf B} = 0$.
Here, ${\textbf v}$ is the velocity of bulk fluid motion,
${\textbf B}$ the magnetic field, ${\textbf J} = \nabla \times
{\textbf B}$ the magnetic current, $\nu$ the kinematic viscosity,
$\eta$ the resistivity. The thermal pressure of the fluid, $p$, is
not an independent variable since, taking the divergence of
Eq.~(\ref{Eq1}), it can be expressed as a function of ${\textbf
B}$ and ${\textbf v}$.
\\
In the case of expanding universe in the radiation era, it has
been shown that the MHD equations are the same as
Eqs.~(\ref{Eq1})-(\ref{Eq2}) provided that time, coordinates, and
dynamical variables are replaced by the following quantities (see,
e.g., Ref.~\cite{Banerjee}):
$t \rightarrow \tilde{t} = \int \! a^{-1 }dt$, ${\textbf x}
\rightarrow \tilde{{\textbf x}} = a {\textbf x}$, ${\textbf B}
\rightarrow \tilde{{\textbf B}} = a^2 {\textbf B}$, $\nu
\rightarrow \tilde{\nu} = a^{-1 } \nu$, $\eta \rightarrow
\tilde{\eta} = a^{-1 }\eta$,
where $a(t)$ is the expansion parameter. Because of the formal
coincidence of the MHD equations in the expanding and
non-expanding universe, we can study the evolution of magnetic
fields in MHD turbulence in both cases in a similar way. For
definiteness, in this paper we shall consider only the case of
non-expanding universe.
\\
It is useful to define the kinetic and magnetic Reynolds numbers,
${\text{Re}} = vl/\nu$ and ${\text{Re}}_B = vl/\eta$,
where $v$ and $l$ are the typical velocity and length scale of the
fluid motion. Magnetohydrodynamic turbulence occurs when
${\text{Re}} \gg 1$ and ${\text{Re}}_B \gg 1$.
We are interested in the evolution of statistically homogeneous
and isotropic magnetic fields. This means that the two-point
correlation tensor $C_{ij}({\textbf x},{\textbf y}) = \langle
B_i({\textbf x}) B_j({\textbf y}) \rangle$, where $\langle ...
\rangle$ denotes ensemble average, is a function of $|{\textbf
x}-{\textbf y}|$ only and transforms as an $SO(3)$ tensor. In
terms of the Fourier amplitudes of the magnetic field,
$B_i({\textbf k},t) = \int \! d^3 x \; e^{i {\textbf k} \cdot
{\textbf x}} \, B_i({\textbf x},t)$, these conditions translate
into \cite{Monin}:
\vspace{-0.1cm}
\begin{eqnarray}
\label{P1} && \!\!\!\!\!\!\!\!\!\!\!\!\!\!\! \langle B_i({\textbf
k},t) B_j({\textbf p},t) \rangle = [(2\pi)^3\!/2] \, \delta
({\textbf k} + {\textbf p}) \nonumber \\
&& \;\;\;\;\;\;\;\;\;\;\; \times \left[ (\delta_{ij} - \hat{k}_i
\hat{k}_j) S(k,t) + i \varepsilon_{ijk} \hat{k}_k A(k,t) \right]
\!\! ,
\end{eqnarray}
where $\hat{k}_i = k_i/k$, $k = |{\textbf k}|$, and
$\varepsilon_{ijk}$ is the totally antisymmetric tensor.
The functions $S$ and $A$ denote the symmetric and antisymmetric
parts of the correlator. They are related to the magnetic energy
and helicity densities in the volume $V$ through
$E_B(t) = (1/2V) \! \int_V d^{\,3} x \, \langle{\textbf B}^2
\rangle = \int_{0}^{\infty} \! dk \, {\mathcal E}_B(k,t)$ and
$H_B(t) = (1/V) \! \int_V d^{\,3}x \, \langle {\textbf A} \cdot
{\textbf B} \rangle = \int_{0}^{\infty} \! dk \, {\mathcal
H}_B(k,t)$,
where
${\mathcal E}_B = k^2 S/(2\pi)^2$ and ${\mathcal H}_B = k
A/(2\pi^2)$
are the magnetic energy and magnetic helicity density spectra, and
${\textbf A}$ is the vector potential.
The kinetic energy, $E_v(t)$, is defined as the magnetic one with
${\textbf B}$ replaced by ${\textbf v}$.
We remember that for all magnetic field configurations, the
magnetic helicity spectrum must satisfy the ``realizability
condition'' \cite{Biskamp}:
$|{\mathcal H}_B| \leq 2k^{-1} {\mathcal E}_B$.
The magnetic field is said to be ``maximally helical" if, for all
$k$, ${\mathcal H}_B$ is of the same sign and saturates the above
inequality.
Moreover, the magnetic helicity is conserved when $\eta = 0$ since
\cite{Biskamp}
$\partial_tH_B = -(2\eta/V) \! \int_V \! d^{\,3} x \, \langle
{\textbf J} \cdot {\textbf B} \rangle = -2\eta \!
\int_{0}^{\infty} \! dk \, k^2 {\mathcal H}_B$.
The relevant length scale in MHD theory, the so-called magnetic
correlation length, is the characteristic length associated with
the large magnetic energy eddies of turbulence and is defined by:
$\xi_B(t) = E_B^{-1} \int_{0}^{\infty} \! d k k^{-1} \, {\mathcal
E}_B$.
The integral form of the realizability condition takes the form:
$|H_B| \leq 2 \xi_B E_B$.
\\
Since we are interested in the case of large Reynolds numbers, we
neglect the dissipation term in Eq.~(\ref{Eq1}).
Moreover, as in Ref.~\cite{Cornwall}, we quasi-linearize the
Navier-Stokes equation (\ref{Eq1}) neglecting the quadratic term
$({\textbf v} \cdot \nabla){\textbf v}$.~\footnote{If we
decompose, in the spirit of mean-field-theory \cite{Biskamp}, the
velocity field into an (almost uniform) average part and a weak,
small-scale fluctuating part, ${\textbf v} = \overline{{\textbf
v}} + \delta {\textbf v}$ with $|\delta {\textbf v}| \ll
|\overline{{\textbf v}}|$, we have ${\textbf v} \cdot \nabla
{\textbf v} \simeq \overline{{\textbf v}} \cdot \nabla \delta
{\textbf v}$. Comparing this term with the Lorentz force, we get
$|{\textbf v} \cdot \nabla {\textbf v}|/ |{\textbf F}_L| \sim
\Gamma |\delta {\textbf v}|/|\overline{{\textbf v}}|$, where
$\Gamma = E_v/E_B$. Hence, the quasi-linear approximation is valid
as long as the condition $\Gamma \gg |\delta {\textbf v}|
/|\overline{{\textbf v}}|$ is satisfied.}
This corresponds to neglect small scale components of velocity
field and to assume that the Lorentz force, ${\textbf F}_L =
{\textbf J} \times {\textbf B}$, acting on the charged particles
of the fluid ``drives'' the development of turbulence on larger
scales:
$\partial_t {\textbf v} \simeq {\textbf F}_L$.
Although the validity of this approximation can be verified only
by a numerical analysis, its use is justified {\it a posteriori}
since our results, as we will find, are in agreement with a
numerical simulations of full MHD equations performed in
Ref.~\cite{Banerjee}.
Finally, we make the common approximation,
\vspace{-0.1cm}
\begin{equation}
\label{v} {\textbf v} \simeq \tau_d {\textbf F}_L,
\end{equation}
where the ``drag time'' $\tau_d$ is the fluid-response time to the
Lorentz force introduced by Sigl in Ref.~\cite{Sigl}.
We note that $\tau_d$ and the characteristic time associated with
kinetic turbulence, the so-called eddy turnover time $\tau_{\rm
eddy} = l/v$, are related by
$\tau_d \simeq \Gamma \tau_{\rm eddy}$,
since
$1 \simeq |{\textbf v}|/|\tau_d {\textbf F}_L| \simeq \Gamma
\tau_{\rm eddy}/\tau_d$.
Here, $\Gamma$ is the ratio of the kinetic and magnetic energy.
What is observed in numerical simulation of MHD equations
\cite{Banerjee} is that, in the non-helical case, turbulence
proceed toward a state of equipartition between magnetic and
kinetic energies ($\Gamma \simeq 1$) while, in the helical case,
though there is no evidence of equipartition, the ratio $\Gamma$
approaches asymptotically to a constant value.
\\
Inserting the above expression for ${\textbf v}$ into the
induction equation (\ref{Eq2}), we get in Fourier space:
\vspace{-0.1cm}
\begin{eqnarray}
\label{Eq16} \!\!\!\!\!\!\!\!\! && (\partial_t + \eta k^2)
B_i({\textbf k}) = \tau_d \! \int \!\! \frac{d^3p}{(2\pi)^3} \!
\int \!\! \frac{d^3q}{(2\pi)^3} \, \varepsilon_{ijk} k_j q_r
B_s({\textbf q}) \times \nonumber \\
\!\!\!\!\!\!\!\!\! && \left[ \varepsilon_{krs} B_n({\textbf
p}\!-\!{\textbf q}) B_n({\textbf k}\!-\!{\textbf p}) \!-\!
\varepsilon_{rsm} B_k({\textbf k}\!-\!{\textbf p}) B_m({\textbf
p}\!-\!{\textbf q}) \right] \! ,
\end{eqnarray}
where summation over repeated indexes is understood. We will work
in ``quasi-normal approximation'' and suppose that the four-point
correlator can be decomposed, in terms of two-point correlator, as
\cite{Biskamp}:
\vspace{-0.1cm}
\begin{eqnarray}
\label{Wick} && \!\!\!\!\!\!\!\!\!\!\!\!\!\!\! \langle
B_i({\textbf k}) B_j({\textbf p}) B_k({\textbf q}) B_l({\textbf
r}) \rangle \nonumber \\
&& \;\;\;\;\;\;\;\;\;\;\;\;\;\;\;\;\;\;\;\; = \langle B_i({\textbf
k}) B_j({\textbf p}) \rangle \langle B_k({\textbf
q}) B_l({\textbf r}) \rangle \nonumber \\
&& \;\;\;\;\;\;\;\;\;\;\;\;\;\;\;\;\;\;\;\; + \; \langle
B_i({\textbf k}) B_k({\textbf q}) \rangle \langle B_j({\textbf p})
B_l({\textbf
r}) \rangle \nonumber \\
&& \;\;\;\;\;\;\;\;\;\;\;\;\;\;\;\;\;\;\;\; + \; \langle
B_i({\textbf k}) B_l({\textbf r}) \rangle \langle B_j({\textbf p})
B_k({\textbf q}) \rangle.
\end{eqnarray}
Multiplying Eq.~(\ref{Eq16}) respectively by $B_i({\textbf k})$
and $A_i({\textbf k})$, and then averaging out we arrive at the
following equations for the magnetic energy and helicity spectra:
\vspace{-0.1cm}
\begin{eqnarray}
\label{Eqspectrum1} \partial_t {\mathcal E}_B \!\!& = &\!\! -2
\eta_{\rm eff} k^2 {\mathcal
E}_B + \alpha_B k^2 {\mathcal H}_B, \\
\label{Eqspectrum2} \partial_t {\mathcal H}_B \!\!& = &\!\! -2
\eta_{\rm eff} k^2 {\mathcal H}_B + 4 \alpha_B {\mathcal E}_B,
\end{eqnarray}
where we have introduced
$\eta_{\rm eff}(t) = \eta + 4 E_B \tau_d/3$ and
$\alpha_B(t) = - \dot{H}_B \tau_d / (3\eta)$.
For simplicity, we will restrict our analysis to magnetic fields
with initial ``fractional helicity'':
${\mathcal H}_B (k,0) = h_B {\mathcal H}^{\mbox{\scriptsize
max}}_B (k,0)$,
where $0 \leq h_B \leq 1$ is the fraction of the initial maximal
helicity
${\mathcal H}^{\mbox{\scriptsize max}}_B (k,t) = 2 k^{-1}
{\mathcal E}_B (k,t)$.
In this case, the solution of
Eqs.~(\ref{Eqspectrum1})-(\ref{Eqspectrum2}) is:
\vspace{-0.2cm}
\begin{eqnarray}
\label{MF13} {\mathcal E}_B (k,t) \!\!& = &\!\! {\mathcal E}_B
(k,0) \exp(-2k^2 \ell_{\rm diss}^2) \nonumber \\
\!\!& \times &\!\! \left[ \cosh(2k \ell_{\alpha}) + h_B \sinh(2k
\ell_{\alpha}) \right] \! ,
\\
\label{MF14} {\mathcal H}_B (k,t) \!\!& = &\!\! {\mathcal
H}^{\mbox{\scriptsize max}}_B (k,0) \exp(-2k^2 \ell_{\rm diss}^2)
\nonumber \\
\!\!& \times &\!\! \left[ \sinh(2k \ell_{\alpha}) + h_B \cosh(2k
\ell_{\alpha}) \right] \! ,
\end{eqnarray}
where we have defined the ``dissipation'' and ``alpha'' lengths,
$\ell_{\rm diss}^2(t) = \int_{0}^{t} \! dt \, \eta_{\rm eff}$ and
$\ell_{\alpha}(t) = \int_{0}^{t} \! dt \, \alpha_B$.
From Eqs.~(\ref{MF13})-(\ref{MF14}) we immediately get that
magnetic fields with maximal initial helicity, $h_B=1$, remain
maximally helical for all times:
${\mathcal H}_B = 2 k^{-1} {\mathcal E}_B$.
To proceed further, we assume that the initial magnetic energy
spectrum can be represented by the following simple function:
${\mathcal E}_B (k,0) = \lambda_B k^p \exp (-2k^2 \ell_{B}^2)$,
where $\lambda_B$ and $\ell_B$ are constants. For $k \ll
\ell_{B}^{-1}$, the magnetic energy spectrum possesses a power law
behavior, while for large $k$ it is suppressed exponentially in
order to have finite energy. The exponential cut-off, $\ell_B$, is
related to the initial correlation length by $\ell_B = \xi_B(0) /
\zeta_B$, where $\zeta_B = \sqrt{2} \, \Gamma(p/2) /
\Gamma[(1+p)/2]$ and $\Gamma(x)$ is the Euler gamma function.
In Ref.~\cite{Caprini}, it was shown that analyticity of the
correlator $C_{ij}({\textbf x},{\textbf y})$ defined on a compact
support forces the spectral index $p$ to be even and equal or
larger than 4.
Now, inserting Eqs.~(\ref{MF13})-(\ref{MF14}) into the expressions
for the magnetic energy and helicity we find:
\\
\vspace{-0.1cm}
\begin{eqnarray}
\label{MF17} \!\!\!\!\!\!\!\!\!\! \frac{E_B(t)}{E_B(0)} \!\!& =
&\!\! (1 + \zeta_{\rm diss}^2)^{-(1+p)/2} \left[ _1 F_1 \!
\left( \frac{1+p}{2} , \frac{1}{2} ; \frac{\chi^2}{2} \right) \right. \nonumber \\
\!\!& + &\!\! \left. p \, h_B \frac{\zeta_B}{2} \, \chi \; _1 F_1
\! \left( \frac{2+p}{2} , \frac{3}{2} ; \frac{\chi^2}{2} \right)
\! \right] \! ,
\\
\label{MF18} \!\!\!\!\!\!\!\!\!\! \frac{H_B(t)}{H_B(0)} \!\!& =
&\!\! (1 + \zeta_{\rm diss}^2)^{-p/2} \left[ _1 F_1 \! \left(
\frac{p}{2} ,
\frac{1}{2} ; \frac{\chi^2}{2} \right) \right. \nonumber \\
\!\!& + &\!\!  \left. \frac{1}{h_B} \frac{2}{\zeta_B} \, \chi \;
_1 F_1 \! \left( \frac{1+p}{2} , \frac{3}{2} ; \frac{\chi^2}{2}
\right) \! \right] \! ,
\end{eqnarray}
where $_1 F_1 (a,b;z)$ is the Kummer confluent hypergeometric
function, and we have defined
$\zeta_{\rm diss} = \ell_{\rm diss}/\ell_B$, $\zeta_\alpha =
\ell_{\alpha}/\ell_B$, and $\chi = \zeta_\alpha /(1 + \zeta_{\rm
diss}^2)^{1/2}$.
Equations (\ref{MF17})-(\ref{MF18}) are integral equations for the
magnetic energy and helicity. They can be solved once the explicit
expression for the drag time is given. This can be done if we
consider the scaling properties of the induction equation. It is
well-known that the full MHD equations (neglecting dissipative
terms) are invariant under the scaling transformations
${\textbf x} \rightarrow \ell {\textbf x}$,
$t \rightarrow \ell^{1-r} t$,
${\textbf v} \rightarrow  \ell^{\,r} {\textbf v}$,
${\textbf B} \rightarrow  \ell^{\,r} {\textbf B}$,
where $\ell > 0$ is the ``scaling factor'' and $r$ is an arbitrary
real parameter \cite{Olesen}. Now, imposing that also the
``reduced'' MHD equations (\ref{Eq2}) and (\ref{v}) are invariant
under these scaling transformations, we get that the drag time is
linear in time. Taking into account the relation between $\tau_d$
and $\tau_{\rm eddy}$ previously discussed, we also have that the
eddy turnover time is asymptotically linear in time. This allow us
to write the drag time as $\tau_d(t) \simeq \Gamma (0) [\tau_{\rm
eddy}(0) + \gamma t]$, where $\gamma = [\Gamma(\infty)/\Gamma(0)]
\lim_{t \rightarrow \infty} \tau_{\rm eddy}(t)/t$ is a constant,
whose explicit value is inessential for the following discussion.
\\
It is useful to define accurately the magnetic Reynolds number and
the eddy turnover time:
${\text{Re}}_B = v_{\rm rms} \xi_B/\eta$, and
$\tau_{\rm eddy} = \xi_B/v_{\rm rms}$,
where, as typical length scale and velocity, we used the magnetic
correlation length and the root-mean-square value of the velocity
field,
$v_{\textmd{rms}}^2 = (1/V) \! \int_V \! d^{\,3} x \,
\langle{\textbf v}^2 \rangle = 2E_v$.
With the aid of the above definitions and introducing the
normalized time $\tau = t/\tau_{\rm eddy}(0)$, the integral
equations (\ref{MF17}) and (\ref{MF18}) can be transformed into
the differential equations
\vspace{-0.341cm}
\begin{eqnarray}
\label{differential1} &&  \frac{d\zeta_{\rm diss}^2}{d\tau} =
\frac{\zeta_B^2}{\mbox{Re}_B(0)} + \frac{2}{3} \, \zeta_B^2 (1 +
\gamma \tau)
\frac{E_B(\tau)}{E_B(0)} \, , \\
\label{differential2} &&  \frac{d\zeta_{\alpha}}{d\tau} =
-\frac{1}{3} \, \zeta_B h_B \mbox{Re}_B(0) (1 + \gamma \tau)
\frac{d}{d\tau} \frac{H_B(\tau)}{H_B(0)} \, ,
\end{eqnarray}
where $E_B$ and $H_B$, as a function of $\zeta_{\rm diss}$ and
$\zeta_{\alpha}$, are given by Eqs.~(\ref{MF17}) and (\ref{MF18}).
For large magnetic Reynolds numbers, the first term in the
left-hand-side of Eq.~(\ref{differential1}) can be neglected with
respect to the second one.

In the non-helical case, $h_B=0$, the solution of
Eqs.~(\ref{differential1})-(\ref{differential2}) is $\zeta_\alpha
= 0$, that is $H_B(t) = 0$ for all times, and
$\zeta_{\rm diss}^2 = [ 1 + \kappa_{\rm diss} ( 2\tau +
\tau^2)]^{2/(3+p)} - 1$,
where $\kappa_{\rm diss} = \gamma (3+p) \zeta_B^2/6$. This, in
turn, gives for $\tau \gg 1$:
\vspace{-0.351cm}
\begin{eqnarray}
\label{Es} && \!\!\!\!\!\!\!\!\!\!\!\!\! E_B(\tau) \simeq \kappa_E
E_B(0) \, \tau^{-2(1+p)/(3+p)},
\\
\label{xis} && \!\!\!\!\!\!\!\!\!\!\!\!\! \xi_B(\tau) \simeq
\kappa_\xi \xi_B(0) \tau^{2/(3+p)},
\end{eqnarray}
where $\kappa_E = \kappa_{\rm diss}^{-(1+p)/(3+p)}$ and
$\kappa_\xi = \kappa_{\rm diss}^{1/(1+p)}$.
\\
It is interesting to observe that, starting from self-similarity
of MHD equations, Olesen obtained the following expression for the
magnetic energy spectrum \cite{Olesen}:
${\mathcal E}_B(k,t) = \lambda_B k^p \, \psi_B(k \,
t^{\frac{2}{3+p}})$,
where $\lambda_B$ is a constant, $\psi_B$ is an unknown
scaling-invariant function, and $p$ is the power-law exponent of
the initial magnetic energy spectrum. Our approach to MHD
equations fixes the expression of the scaling-invariant function
to
$\psi_B (x) = \exp [-2 (x/x_s)^2]$,
with
$x_s = \kappa_\xi^{-1} \ell_B^{-1} [\tau_{\rm eddy}(0)]^{2/(3+p)}$.
%
%
\begin{figure}[t]
\begin{center}
\includegraphics[clip,width=0.381\textwidth]{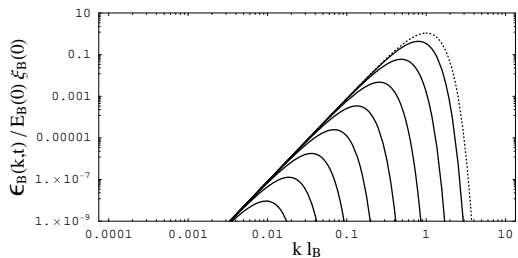}
\caption{Magnetic energy spectrum in the non-helical case for
$p=4$, with $\gamma = 1$. The dotted line corresponds to the
initial spectrum, while continuous lines correspond, from left to
right, to $t/\tau_{\rm eddy}(0) = 1,10,10^2,...,10^7$.}
\end{center}
\end{figure}
%
\\
In Fig.~1, we plot the spectrum of the magnetic energy for the
case $p=4$ at different times. It is clear that, in the
non-helical case, the decay of the magnetic field and the growth
of the correlation length proceed through the so-called {\it
selective decay} discussed by Son in Ref.~\cite{Son}: there is no
direct transfer of magnetic energy from small scales (large
wavenumbers) to large scales (small wavenumbers) but, simply,
modes with larger wavenumbers decay faster than those whose
wavenumbers are small. Consequently, as the turbulence operates,
the magnetic field survives only on larger and larger scales.
\\
In the helical case, the evolution of the system goes through two
different regimes depending on the value of $\chi(t)$ which is an
increasing function of time. Until when $\chi \ll 1$ the system
behaves as if the magnetic helicity were zero: the system evolves
by selective decay and, consequently, the asymptotic solutions are
the same obtained previously. Afterwards, when $\chi \gg 1$, the
system enters and persists in a phase characterized by a transfer
of magnetic energy from small to large scales, a mechanism known
as {\it inverse cascade} \cite{Biskamp}.
The asymptotic ($\tau \rightarrow \infty$) solutions in this
latter phase are:
$\zeta_{\rm diss}(\tau) \simeq c_{\rm diss} (\ln \tau)^{1/6}
\tau^{2/3}$ and
$\zeta_{\alpha}(\tau) \simeq c_{\alpha} (\ln \tau)^{2/3}
\tau^{2/3}$,
where $c_{\rm diss} = (\gamma^2 p/12)^{1/6} \zeta_B h_B^{1/3}$ and
$c_{\alpha} = (4p/3)^{1/2} c_{\rm diss}$.
Consequently, we have:
\vspace{-0.25cm}
\begin{eqnarray}
\label{expansion-Energy} E_B(\tau) \!\!& \simeq &\!\! c_E E_B(0)
 \, (\ln \tau)^{1/3} \, \tau^{-2/3},
\\
\label{expansion-correlation} \xi_B(\tau) \!\!& \simeq &\!\! c_\xi
\xi_B(0) \, (\ln \tau)^{-1/3} \, \tau^{2/3},
\end{eqnarray}
with
$c_E = (2p/3\gamma)^{1/3} h_B^{2/3}$, and
$c_\xi = h_B c_E^{-1}$.
From the above equations, we directly obtain the relation $E_B
\xi_B\simeq H_B/2$. This means that a magnetic field with initial
fractional helicity becomes maximally helical approximatively
after the system enters into the inverse-cascade regime. More
accurately, we can find the time when this happens, $\tau_h$,
matching the product of asymptotic solutions
(\ref{Es})-(\ref{xis}) and
(\ref{expansion-Energy})-(\ref{expansion-correlation}). It
results:
$\tau_{h} \simeq \kappa_{\rm diss}^{-1/2} \, h_B^{-(3+p)/2p}$.
In Fig.~2 we present the result of a numerical integration of
Eqs.~(\ref{differential1})-(\ref{differential2}) for
${\text{Re}}_B(0) = 10^{15}$, $p=4$, and $h=10^{-3}$. It is
evident from the figure that the analytical expansions
[non-helical solution for $\tau \lesssim \tau_h$ and
Eqs.~(\ref{expansion-Energy})-(\ref{expansion-correlation}) for
$\tau \gtrsim \tau_h$] fit very well the numerical solution.
Because of quasi-conservation of magnetic helicity, small-scale
modes are not dissipated during the decay but their energy is
transferred to larger scales: this process of inverse cascade is
manifest in the magnetic energy spectrum shown in Fig.~2.
%
%
\begin{figure}[t]
\begin{center}
\includegraphics[clip,width=0.381\textwidth]{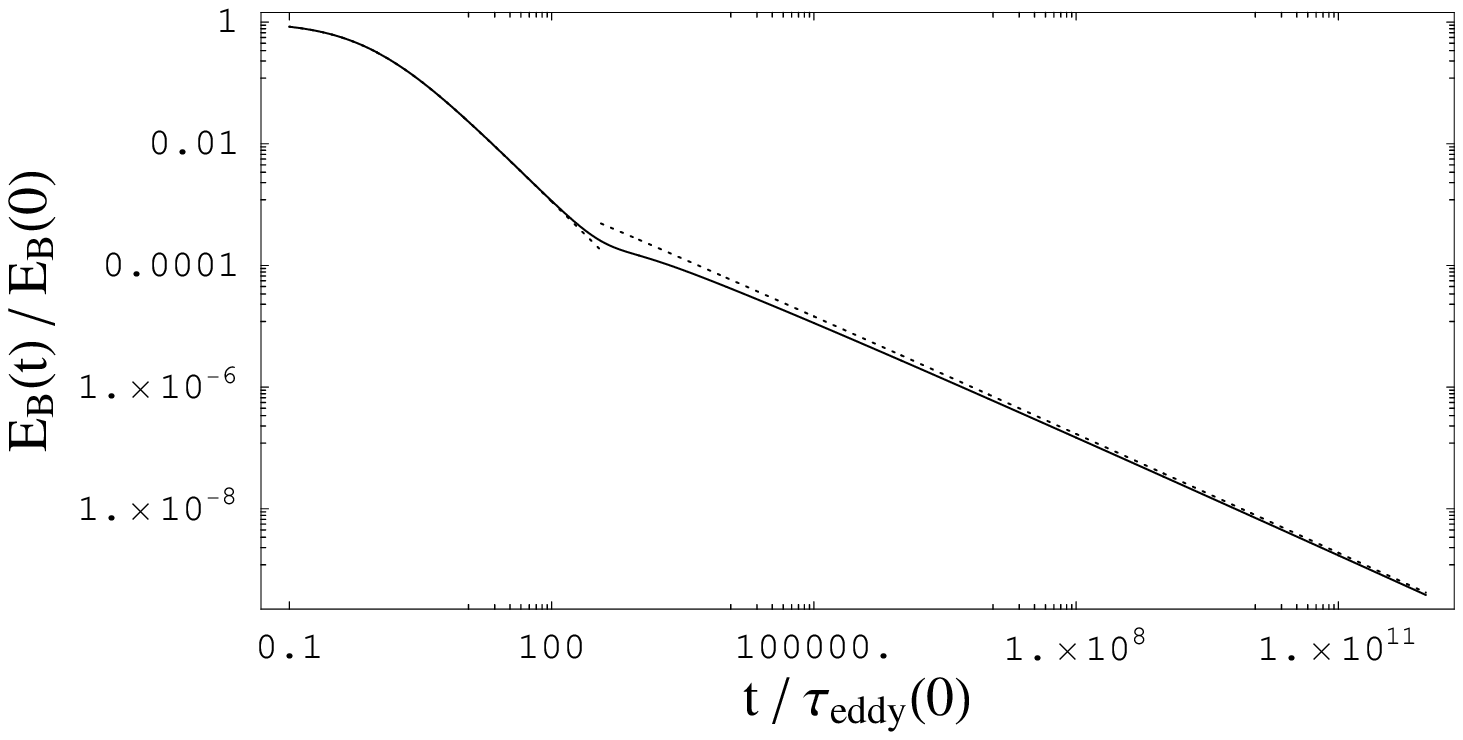}
\\
\hspace{0.01cm}
\includegraphics[clip,width=0.373\textwidth]{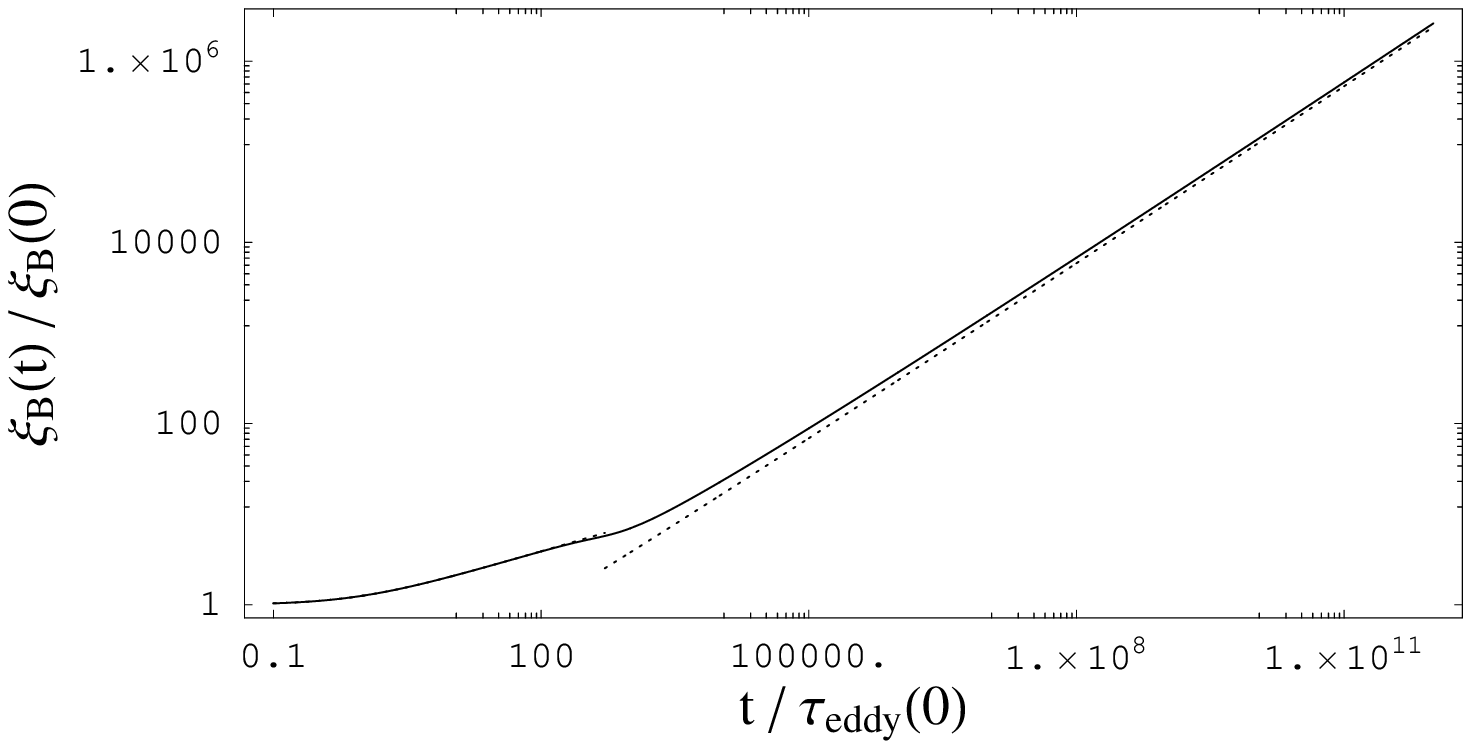}
\\
\includegraphics[clip,width=0.381\textwidth]{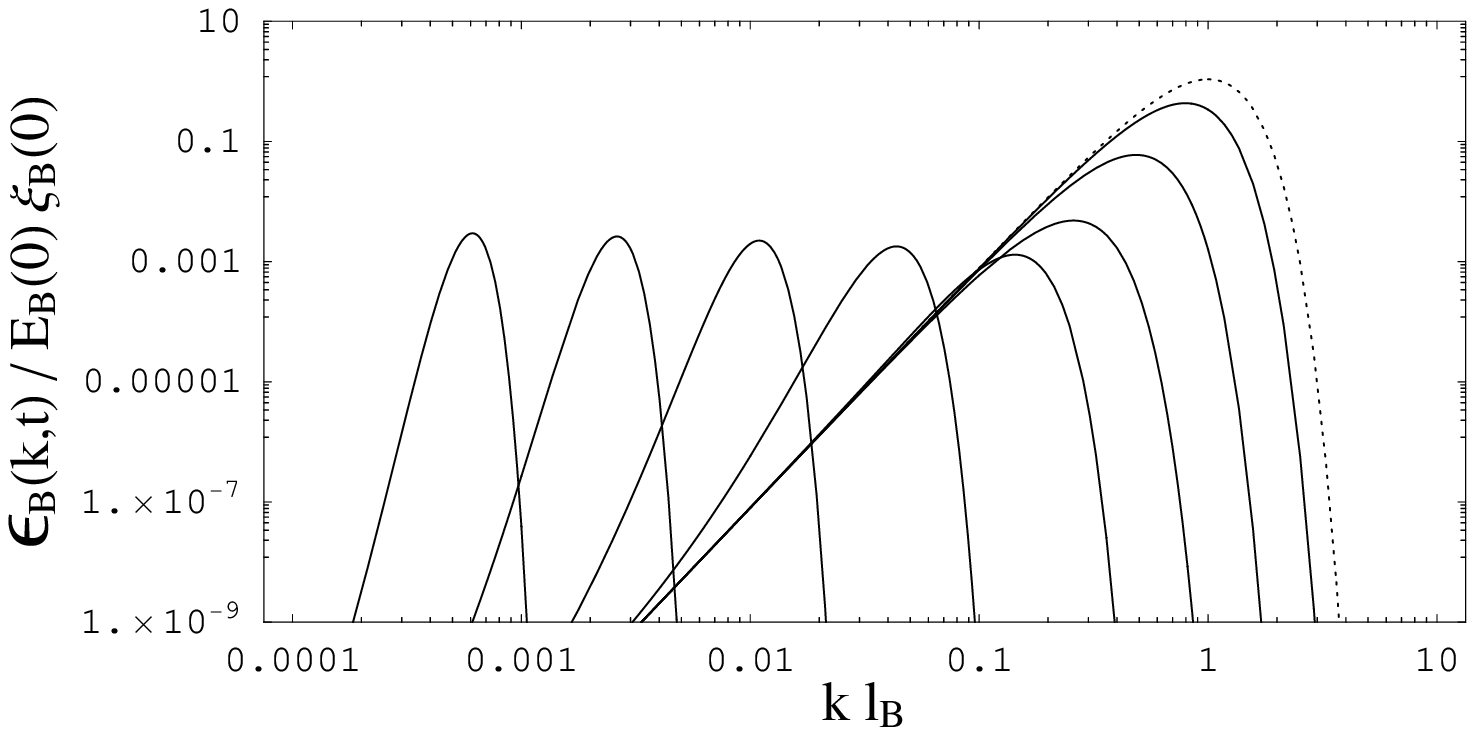}
\caption{Result of a numerical integration of
Eqs.~(\ref{differential1})-(\ref{differential2}) for
${\text{Re}}_B(0) = 10^{15}$, $p=4$, $h=10^{-3}$, with $\gamma =
1$.
Upper panel: magnetic energy; middle panel: correlation length;
Dotted lines correspond to analytical expansions. Lower panel:
magnetic energy spectrum; the dotted line corresponds to the
initial spectrum, while continuous lines correspond, from left to
right, to $t/\tau_{\rm eddy}(0) = 1,10,10^2,...,10^7$.}
\end{center}
\end{figure}
%
\\
It is worth noting that our final results,
Eqs.~(\ref{Es})-(\ref{expansion-correlation}), apart logarithmic
factors, agree very well both with scaling arguments and results
of a numerical integration of full MHD equations presented in
Ref.~\cite{Banerjee}.

%
\vspace{-0.05cm}
In conclusion, we have studied the evolution of statistically
homogeneous and isotropic magnetic fields in the context of freely
decaying magnetohydrodynamic turbulence. By quasi-linearizing the
Navier-Stokes equation, we have solved analytically the induction
equation in quasi-normal approximation.
We have found that, if the initial magnetic field is not helical,
the evolution of the magnetic field proceeds through selective
decay of magnetic modes: magnetic power on small scales is washed
out by turbulence effects more effectively than on large scales.
During this process, the correlation length grows as $\xi_B
\propto t^{2/(3+p)}$, while the magnetic energy decays in time as
$E_B \propto t^{-2(1+p)/(3+p)}$, where $p$ is the index of the
initial power-law spectrum.
In the helical case, the evolution of the system goes through two
different phases: selective-decay phase in which the system
evolves as if the magnetic helicity were zero and inverse-cascade
phase. The first phase ends when quasi-conservation of magnetic
helicity starts to trigger an inverse cascade of the magnetic
field: small-scale modes are no more completely dissipated during
turbulence but their energy is partially transferred to larger
scales. This causes a faster growth of the correlation length and
a slower dissipation of the magnetic energy with respect to the
non-helical case.
The time when the system enters into the inverse-cascade regime is
proportional to $h_B^{-(3+p)/2p}$ times the initial eddy turnover
time, where $h_B$ is fraction of the maximal initial magnetic
helicity.
Moreover, the process of inverse cascade erases any information
about the initial structure of the magnetic field, so that the
evolution laws of energy and correlation length are $E_B \propto
(\log t)^{1/3} t^{-2/3}$ and $\xi_B \propto (\log t)^{-1/3}
t^{2/3}$, whatever is the value of $p$.
\\
In a cosmological context, these results are of interest when
studying the evolution of primordial magnetic fields before
neutrino decoupling. Indeed, during the period of neutrino (or
photon) free-streaming, as well as after recombination, the
equations governing the evolution of magnetic fields differ from
those studied here \cite{Banerjee} and then our results do not
apply. Nevertheless, our approach to MHD equations can be suitably
extended to these last cases and an appropriate analysis is in
progress.
%
\vspace{-0.65cm}

\end{document}